\documentclass[11pt,twoside]{article}

\usepackage{asp2004}
\usepackage{epsf}
\usepackage{psfig}
\usepackage{lscape}

\begin{document}
\title{HST and FUSE Spectroscopy of the DAO-type Central Star LS\,V+4621}
\author{T\@. Rauch, M\@. Ziegler, K\@. Werner}
\affil{Institut f\"ur Astronomie und Astrophysik, Universit\"at T\"ubingen, Sand 1, 72076 T\"ubingen, Germany}
\author{J.W\@. Kruk}
\affil{Department of Physics and Astronomy, Johns Hopkins University, Baltimore, MD 21218, U.S.A.}

\begin{abstract}
The DAO-type white dwarf LS\,V+4621 is the hydrogen-rich central star of the possible planetary nebula Sh\,2-216.
We have taken high-resolution, high-S/N ultraviolet spectra with STIS aboard the HST and FUSE
in order to constrain its photospheric parameters.
A detailed spectral analysis by means of state-of-the-art NLTE model-atmosphere techniques is presented
which includes the determination the individual abundances of iron-group elements.
\end{abstract}

\section{A brief history of Sh\,2-216 and LS\,V+4621}
\label{sect:history}

Sharpless 2-216 (Sh\,2-216) has been discovered as a ``curious emission-line nebula'' \citep*{f1981}.
At a distance of $d=130\,\mathrm{pc}$ \citep{h1997}, it is the closest possible planetary nebula (PN) known 
with an apparent diameter of $100'$. Sh\,2-216 has experienced a mild interaction with the 
interstellar medium \citep*[ISM;][]{tmn1995}. From its distance and proper motion, 
\citet{kea2004} have determined that it has a thin-disk orbit of low inclination and eccentricity.

LS\,V+4621 has been identified as the exciting star of Sh\,2-216 by proper motion measurements \citep{cr1985}.
\citet{tn1992} have demonstrated that LS\,V+4621, which is the brightest ($m_\mathrm{V} = 12.67$) DAO white dwarf 
(WD\,0439+466) known,
has the properties ($T_\mathrm{eff}=90\,\mathrm{kK}$, $\log g = 7$ (cgs), $\mathrm{He/H} = 0.01$ by number)
to ionize the surrounding nebula.

\section{On the Balmer-line problem in LS\,V+4621}
\label{sect:blp}

\citet{n1992, n1993}, \citet{ns1993}, and \citet{nr1994} have reported that a problem (BLP) exists
to reproduce lines of the Balmer series (with NLTE model atmospheres) 
simultaneously at a given $T_\mathrm{eff}$ in hot DA white dwarfs.
\citet{brl1993} found (in LTE calculations for DAO WDs) that the BLP is reduced by the consideration of 
metal-line blanketing, and \citet{bea1994} could show clearly that the presence of heavy metals is the source of the BLP --
in other words, pure hydrogen models are not well suited for the spectral analysis of hot DA WDs in general.

\citet{w1996} calculated NLTE model atmospheres for LS\,V+4621 based on parameters of \citet{tn1992}
and introduced C, N, and O (at solar abundances) in addition. Surface cooling by these metals as well as the detailed
consideration of the Stark line broadening in the model-atmosphere calculation has the effect that the
BLP almost vanishes in LS\,V+4621. Later, \citet{kw1998} could demonstrate that these model atmospheres
reproduce well HUT (Hopkins Ultraviolet Telescope) observations of LS\,V+4621 within $912 - 1840\,\mathrm{\AA}$
at $T_\mathrm{eff}=85\,\mathrm{kK}$ and $\log g = 6.9$.

\section{UV observations}
\label{sect:obs}

Spectral analysis by model-atmosphere techniques needs observations
of lines of subsequent ionization stages in order to evaluate the ionization equilibrium (of a particular
species) which is a sensitive indicator of $T_\mathrm{eff}$. Since stars with $T_\mathrm{eff}$ as high as 
$\approx 90\,\mathrm{kK}$ have their flux maximum in the EUV wavelength range and due to the high degree
of ionization, most of the metal lines are found in the UV range. Thus, high-S/N and high-resolution UV spectra
are a prerequisite for a precise analysis. Consequently, we employed HST/STIS (Space Telescope Imaging Spectrograph
aboard the Hubble Space Telescope)
and FUSE (Far Ultraviolet Spectroscopic Explorer) in order to obtain suitable data.

A STIS spectrum ($1144 - 1729\,\mathrm{\AA}$, exposure time 5.5\,ksec, resolution = 0.06\,\AA) was taken in
2000 and was processed by the standard pipeline data reduction. A FUSE observation ($905 - 1195\,\mathrm{\AA}$,
67.6\,ksec, 0.05\,\AA) was performed in 2003/2004 and reduced by J.W.K\@. (Feb 2005).

\section{Evolutionary models}
\label{sect:evo}

For the evolution of LS\,V+4621 standard evolutionary models for hydrogen-rich post-AGB stars,
e.g\@. \citet{s1983} and \citet{bs1990} are appropriate. 
Recently, new evolutionary calculations for DA WDs with a thin hydrogen envelope have been 
presented by \citet{aea2005}. These evolutionary calculations are used to compare with and to determine 
stellar masses, luminosities, and post-AGB ages \citep[see][for details]{zea2007}.

\section{Spectral analysis}
\label{sect:analysis}

The plane-parallel, static, and chemically homogeneous models 
used in this analysis are calculated with {\sc TMAP}, the T\"ubingen NLTE Model Atmosphere Package
\citep*{wea2003}. H+He+C+N+O+Mg+Si are considered with ``classical'' model atoms \citep[cf\@.][]{r2003}.
For Ca+Sc+Ti+V+Cr+Mn+Fe+Co+Ni individual model atoms are constructed by
{\sc IrOnIc} \citep*{rd2003}, using a statistical approach in order to treat the
overwhelmingly large number of atomic levels and line transitions by the introduction
of ``super-levels'' and ``super-lines''. In total 531 levels are treated in NLTE,
combined with 1761 individual lines and about 9 million iron-group lines, taken
from \citet{k1996} as well as from the OPACITY and IRON projects \citep{sea1994,hea1993}.

\begin{figure}[ht]
\epsfxsize=\textwidth
\epsffile{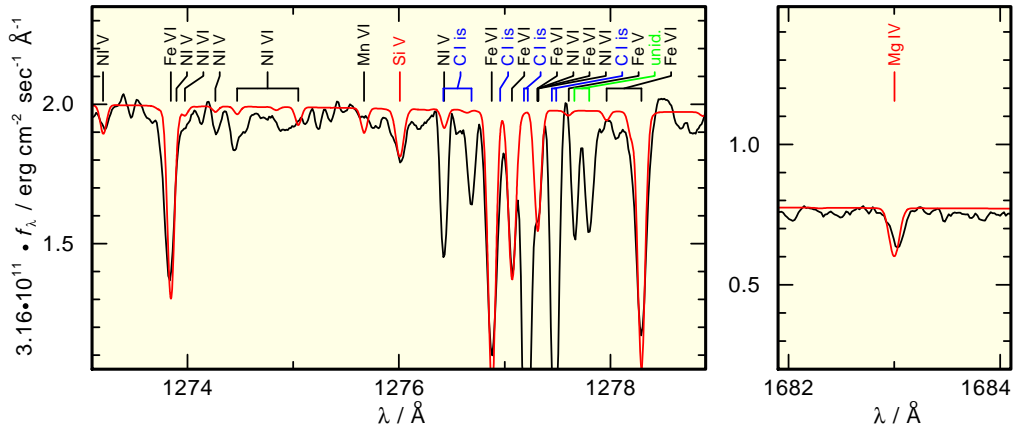}
\caption{Examples for the identification of \ion{Si}{v} and \ion{Mg}{iv} lines in the STIS spectrum of LS\,V+4621.
         ``is'' denotes interstellar, ``unid\@.'' means unidentified.}
\label{fig:ident}
\end{figure}

\subsection{The STIS spectrum}
\label{subsect:STIS}

\begin{figure}[ht]
\epsfxsize=\textwidth
\epsffile{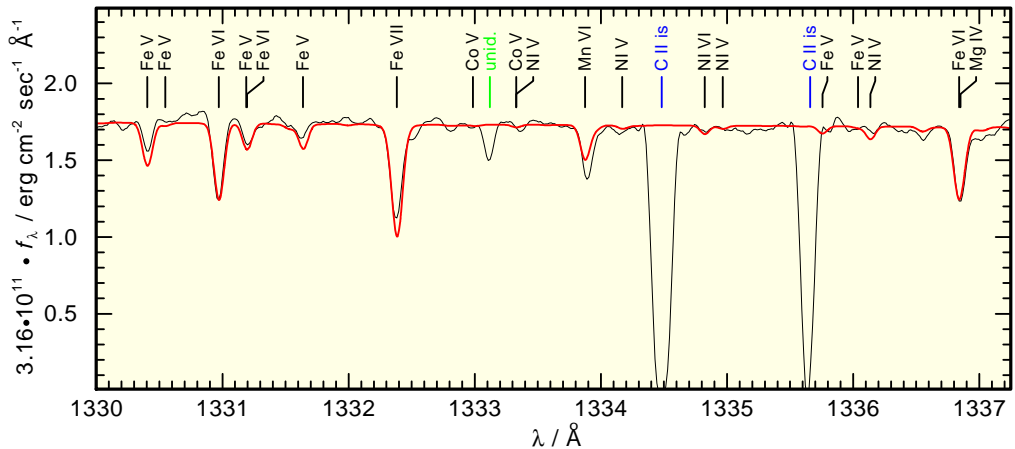}
\caption{Section of the STIS spectrum of LS\,V+4621 compared with our final model. Identified lines are marked.
Note the \ion{Mn}{vi} $\lambda = 1333.87\,\mathrm{\AA}$ line.}
\label{fig:teff}
\end{figure}

The STIS spectrum has a very good S/N ($>50$, Fig.~\ref{fig:ident}) 
and allows to identify about 95\,\% of all spectral features
(photospheric as well as interstellar absorptions). We were able to identify some \ion{Si}{v} lines
(Fig.~\ref{fig:ident}), which allow to use the \ion{Si}{iv}\,/\,\ion{Si}{v} ionization equilibrium
(Sect.~\ref{sect:obs}) for our analysis \citep[cf\@.][]{jea2007}. 
Moreover, we could identify \ion{Mg}{iv} lines (Fig.~\ref{fig:ident}) -- 
up to our knowledge -- for the first time in these objects \citep[cf\@.][]{zea2007}.

From a detailed comparison of \ion{H}{i} Ly\,$\alpha$ with the observation, we determined
$N_\mathrm{H\,I} = 8.5 \pm 0.1 \cdot 10^{19} \mathrm{cm^{-2}}$. Given this value, the Galactic
reddening law of \citet{gl1989} yields $E_\mathrm{B-V} = 0.021$. We achieve the best match to the
continuum slope with $E_\mathrm{B-V} = 0.065 \pm 0.04$. This is in agreement with the result of
$E_\mathrm{B-V} = 0.1$ that \citet{kw1998} derived from the analysis of a HUT spectrum. 
We determined a photospheric radial velocity of $v_\mathrm{rad}  = 20.4 \pm 0.4\,\mathrm{km\,sec^{-1}}$. This is 
significantly higher than the values of $11.9\,\mathrm{km\,sec^{-1}}$ and $11.1\,\mathrm{km\,sec^{-1}}$
  measured from IUE (International Ultraviolet Explorer) spectra         
by \citet*{tn1992} and \citet*{hea1998}, respectively.
Such a large difference is possible if the object has not been located in the middle of the
IUE aperture (Holberg priv\@. comm.). The higher $v_\mathrm{rad}$ will have some influence on the
calculation of the Galactic orbit of LS\,V+4621 (Sect.~\ref{sect:history}) and hence, investigation
on the interaction of Sh\,2-216 with the ISM.

As an example for the spectral analysis, we have selected a section between 1330 and 1137\,\AA\ in order to demonstrate
that we can fit \ion{Fe}{v}, \ion{Fe}{vi}, and \ion{Fe}{vii} lines simultaneously (Fig.~\ref{fig:teff}). 
Because of the sensitivity of the ionization balance, this allows to determine $T_\mathrm{eff}$ within very small error limits.

\subsection{The FUSE spectrum}
\label{subsect:FUSE}

\begin{figure}[ht]
\epsfxsize=\textwidth
\epsffile{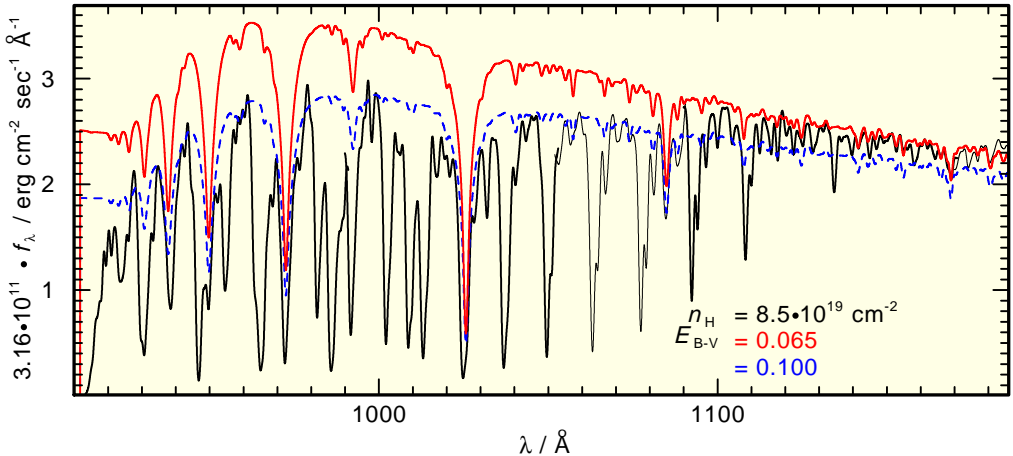}
\caption{FUSE spectrum of LS\,V+4621 compared with our final model at two different reddenings. For clarity, observation
and model fluxes are smoothed with a Gaussian of 1\,\AA\ (FWHM).
}
\label{fig:FUSE}
\end{figure}

The FUSE spectrum of LS\,V+4621 is heavily contaminated by interstellar absorption features (Fig.~\ref{fig:FUSE}).
Our final model fits to the observation at wavelengths $< 1090\,\mathrm{\AA}$ better with a higher reddening 
($E_\mathrm{B-V} = 0.1$). 
Some of this difference in continuum slope can be explained by the interstellar H$_2$ opacity,
but an increase in the interstellar extinction appears to be necessary.

The analysis of the FUSE spectrum and an investigation on the interstellar absorption 
is presented in more detail by \citet[][and these proceedings]{zea2007}.

\section{Conclusions}
\label{sect:concl}

\begin{figure}[ht]
\epsfxsize=\textwidth
\epsffile{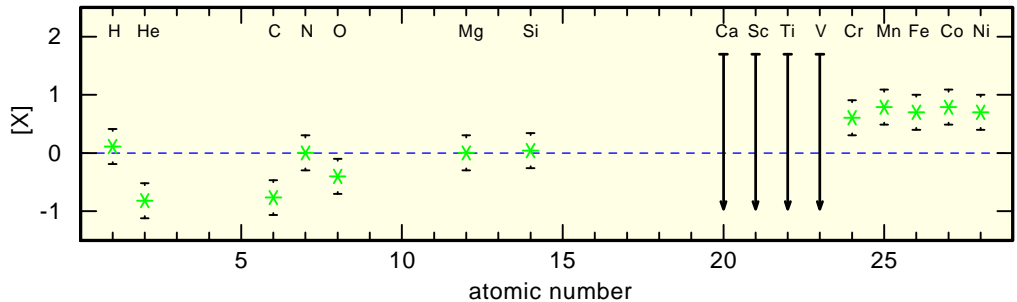}
\caption{Photospheric abundances of LS\,V+4621 determined from detailed line profile fits. 
[x] denotes log abundance / solar abundances of species x.
The relatively large error ranges for Ca, Sc, Ti, and V are due to the fact, that not any line of these
elements could be identified, neither in the STIS nor in the FUSE spectrum (Sect.~\ref{subsect:STIS}).
Thus, we determined upper limits only.}
\label{fig:abundances}
\end{figure}

From the \ion{N}{iv} -- \ion{N}{v}, \ion{O}{iv} -- \ion{O}{v}, \ion{Si}{iv} -- \ion{Si}{v}, and
\ion{Fe}{v} -- \ion{Fe}{vii} ionization equilibria, we were able to determine
$T_\mathrm{eff} = 95 \pm 2\,\mathrm{kK}$ with -- for these objects -- unprecedented precision. 
Since this is a prerequisite for reliable abundance determinations, their error limits 
(Fig.~\ref{fig:abundances}) are also relatively small. 
The formerly determined surface gravity of $\log g = 6.9$ \citet{tea2005} has been confirmed.

The derived abundance pattern (Fig.~\ref{fig:abundances}) gives evidence for an interplay of
gravitational settling (e.g\@. the He and C abundances are strongly decreased by a factor of $\approx 0.15$) 
and radiative levitation (iron-group elements show an up to $\approx 6\times$ solar increased abundance).  

In our metal-line blanketed NLTE model atmospheres which include the opacity of 16 species
from H to Ni (Sect.~\ref{sect:analysis}),
the BLP (Sect.~\ref{sect:blp}) vanishes in the available medium-resolution optical spectra (at $S/N \approx 30$) --
now adequate high-resolution and high-S/N optical spectra are desirable in order to investigate if
there are still remaining problems.

The reddening of $E_\mathrm{B-V} = 0.065 \pm 0.04$ towards LS\,V+4621 is much higher than expected from the
Galactic reddening law (Sect.~\ref{subsect:STIS}) possibly because of additional reddening due to dust in the
nebula Sh\,2-216.

\section{Future work}
\label{sect:future}

State-of-the-art NLTE spectral analysis has arrived at a high level of sophistication. However, it is hampered
by the lack of reliable atomic data for metal lines. We are able to identify/reproduce about 95\% of all spectral
lines in the STIS spectrum of LS\,V+4621 and it is likely that unidentified lines 
(e.g\@. in Figs.~\ref{fig:ident}, \ref{fig:teff})
simply stem from the most prominent ions as well, but their wavelengths are not sufficiently well known.
E.g\@. for \ion{Fe}{vii}, \citet{k1996} provides only 22 laboratory measured (POS) lines and 1952 lines
with theoretical line positions (LIN lines). This situation is even worse for other ions (\ion{Fe}{vi}: 224 and 58664,
respectively) and species.
Thus, it should be a challenge for atomic physics to provide properly measured atomic data (also for
highly ionized elements) which will then strongly improve future spectral analyses.

\acknowledgements T.R\@. was supported by the DLR (grants 50\,OR\,0201 and 05\,AC6VTB). 
M.Z\@. thanks the Astronomische Gesellschaft for a travel grant.
J.W.K\@. is supported by the FUSE project, funded by NASA contract NAS532985.
This research has made use of the SIMBAD Astronomical Database, operated at CDS, Strasbourg, France.

\end{document}